\documentclass[aps,pre,twocolumn,amsmath,amssymb,showpacs,showkeys,floatfix]{revtex4-2}
\usepackage{epsfig}
\usepackage{subcaption}
\usepackage{graphicx}
\usepackage{setspace}
\usepackage{bm}
\usepackage{float}
\usepackage{amsmath}
\usepackage{braket}
\usepackage{xcolor}
\usepackage[colorlinks,bookmarks=false,citecolor=blue,linkcolor=red,urlcolor=blue]{hyperref}

\pacs{physics}

\begin{document}
\title{$J=0$ metastable state of $\mathrm{Th}^{2+}$ for a hyperfine-free nuclear clock}
\author{S. Sagar Maurya}
\author{V. Lal}
\author{J. Tiedau}
\author{M. V. Okhapkin}
\author{E. Peik}
\email {ekkehard.peik@ptb.de}
\affiliation{Physikalisch-Technische Bundesanstalt (PTB), Bundesallee 100, 38116 Braunschweig, Germany}

\begin{abstract}
We present measurements on a metastable state in $\mathrm{Th}^{2+}$ with the electronic configuration  $6d^2\,{}^3P_0\,(5090\ \mathrm{cm^{-1}})$. This is motivated by the prospect of using the state in laser excitation of the low-energy $^{229}$Th nuclear resonance independent from the leading hyperfine interactions. 
The $6d^2\,{}^3P_0$ state has no dipole-allowed radiative decay channel and is connected to the second ground state $6d^2\,{}^3F_2\,(63\ \mathrm{cm^{-1}})$ through an electric quadrupole transition only. We populate the state 
by laser excitation at 484~nm via a higher excited level and detect its population in laser-induced fluorescence.
The isotope shift of the $J=0$ level between $^{232}\mathrm{Th}^{2+}$ and $^{229}\mathrm{Th}^{2+}$ is determined as a measure of the interaction of electronic and nuclear charge distributions. 
The lifetime of the level in our ion trap with buffer gas is limited by collisional mixing with 
the nearby state $5f6d\,{}^3G_3\,(5060\ \mathrm{cm^{-1}})$.
In ultrahigh vacuum, it could serve as a hyperfine-free nuclear clock that is largely immune to field-induced frequency shifts via the electron shell.   
\end{abstract}

\maketitle

\section{Introduction}
The exceptionally low-energy nuclear transition $(8.4\,\mathrm{eV})$ in $^{229}\mathrm{Th}$ represents a unique opportunity for the realization of a nuclear optical clock \cite{E.Peik_2003,PhysRevLett.108.120802}. Direct laser excitation of the thorium nucleus in a doped crystal \cite{PhysRevLett.132.182501,PhysRevLett.133.013201} has resolved the long-standing uncertainty in the transition frequency and yielded a precise value. Furthermore, detailed studies with thorium-doped crystals, including measurements of the nuclear electric quadrupole structure \cite{Zhang2024} and  investigations demonstrating frequency reproducibility \cite{Ooi2026}, have significantly strengthened the foundation for precision nuclear frequency metrology.

In contrast to the solid-state approach, a nuclear clock based on a trapped ion in a radio-frequency trap could offer higher accuracy by isolating the ion from environmental perturbations. However, due to the hyperfine coupling between the nucleus and the valence electrons, the sensitivity of the transition frequency to external fields depends on the electron quantum numbers. Two classes of states have been identified that minimize the sensitivity: (i) States with electronic angular momentum $J=0$ or $J=1/2$ \cite{E.Peik_2003}. 
(ii)  Stretched states with the maximum value of the total angular momentum $F=|m_F|=J+I$ \cite{PhysRevLett.108.120802}. 
For both classes of states, the field-induced systematic frequency shifts can be suppressed so that they become negligible in comparison to the second-order Doppler effect which will dominate the uncertainty budget \cite{E.Peik_2003,PhysRevLett.108.120802, PhysRevLett.130.103201, 88jn-939s}. 

The electronic structures of thorium cations in charge states Th$^+$ to Th$^{4+}$ show significantly different properties.
The level schemes of Th$^+$ and Th$^{2+}$ are characterized by a high density of states and strong configuration mixing \cite{Redman_2014, Biémont_2002, PhysRevA.90.032512, Wyart_1981}.  
Th$^{3+}$, originally proposed for the nuclear clock \cite{E.Peik_2003}, presents a simple level structure that is amenable to laser cooling \cite{PhysRevLett.106.223001, PhysRevLett.102.233004}. The ground state possesses $J=5/2$ and a metastable $s$-state with $J=1/2$ and an approximate lifetime of $600$ ms exists \cite{PhysRevA.86.060501, PhysRevA.74.042511}. Th$^{4+}$ is radon-like with a $J=0$ ground state, but does not possess a resonance line for laser cooling or state detection in the wavelength range of common laser sources \cite{PhysRevA.97.012511,zcsk-3lw5}.     

Experimental studies of the electronic hyperfine structure of the $^{229}\mathrm{Th}$ nuclear ground state have been performed in trapped $^{229}\mathrm{Th}$ ions of different charge states. In particular, laser-cooled $^{229}\mathrm{Th}^{3+}$ ions have been used for high-resolution spectroscopy \cite{PhysRevA.111.L050802,PhysRevLett.106.223001}, while buffer-gas-cooled $^{229}\mathrm{Th}^{+}$ and $^{229}\mathrm{Th}^{2+}$ ions have also been investigated \cite{PhysRevA.92.020503,Thielking2018}. Furthermore, hyperfine-resolved spectroscopy of both the nuclear ground state and the isomeric state of $^{229}\mathrm{Th}$, produced as recoil ions from the $\alpha$-decay of $^{233}$U, has been demonstrated in trapped $^{229}\mathrm{Th}^{2+}$ and $^{229}\mathrm{Th}^{3+}$ ions under buffer-gas cooling conditions \cite{Thielking2018, Yamaguchi2024}. Observation of the isomeric state 
in laser-cooled or sympathetically cooled ions has not yet been achieved. The successful realization of such measurements would enable more precise determinations of the isomer's nuclear moments and charge radius. Similarly, direct laser excitation of the nucleus in trapped ions remains experimentally challenging due to the limited available laser power at 148 nm and the achievable low excitation rate \cite{Beeks2021}. There is an increasing interest in $\mathrm{Th}^{2+}$, which has favorable properties for nuclear excitation and as a potential candidate for a nuclear clock. In particular, it has been proposed as a system for nuclear excitation through electronic bridge processes  \cite{PhysRevA.111.053109} and for two-photon nuclear excitation \cite{Yudin2025}.

In this work, we investigate a metastable electronic level with total angular momentum $J = 0$  in Th$^{2+}$ that is particularly attractive due to its symmetry and the isolation it provides from external perturbations to the nucleus through the hyperfine interaction. It may serve as a hyperfine-free nuclear clock candidate, analogous to a recent proposal for $\mathrm{Th}^{4+}$ clock schemes that exploit the closed-shell ground state configuration with $J=0$ \cite{zcsk-3lw5}. The level, designated as $6d^2\,{}^3P_0\,(5090\ \mathrm{cm^{-1}})$ has no electric dipole-allowed transition to any lower level. 
Here, we show how the state can be populated via laser excitation to higher excited states. 
The population is detected in laser-induced fluorescence, and the isotope shift between $^{232}\mathrm{Th}^{2+}$ and $^{229}\mathrm{Th}^{2+}$ is measured. Furthermore, we also determine its lifetime as determined by collisions in different buffer gases under varying pressures. 
The radiative lifetime of this level has been calculated to be 95.5 s \cite{PhysRevA.90.032512} for decay through the electric quadrupole transition (E2) to the $6d^2\,{}^3F_2\,(63\ \mathrm{cm^{-1}})$ state at a wavelength of 1989~nm.

Due to strong configuration mixing, designations of levels of Th$^{2+}$  in $LS$ coupling can only give an approximate description. In the following, we label the states by their energies (in cm$^{-1}$) and total angular momentum $J$ as a subscript.

\section{Experimental setup}
\begin{figure}[h]
    \centering
    \includegraphics[width=0.48\textwidth]{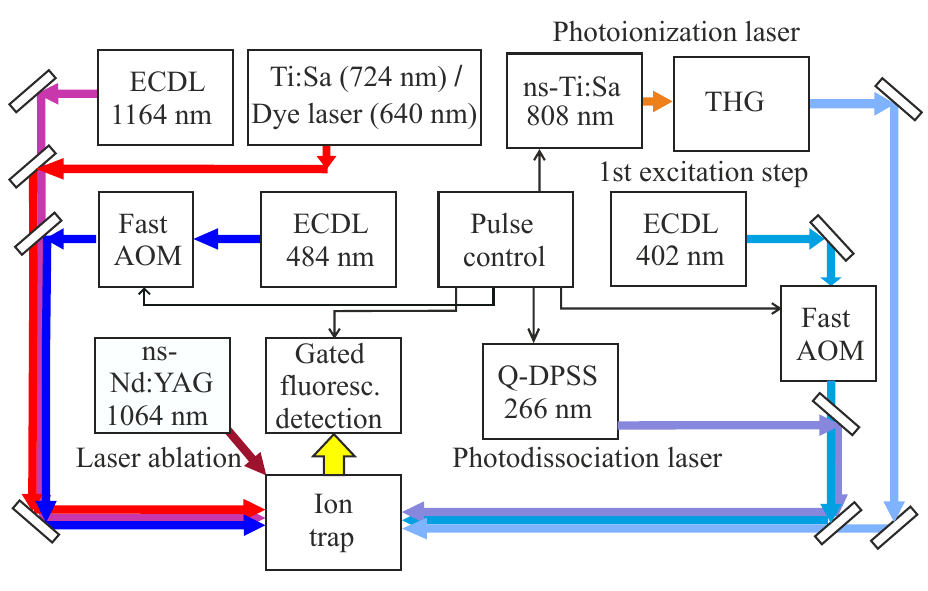}
    \caption{Optical setup for the spectroscopy of $\mathrm{Th}^{2+}$ ions. An external cavity diode laser (ECDL) at 402 nm is used to monitor $\mathrm{Th}^{+}$ ions and, in combination with photoionization laser, to produce $\mathrm{Th}^{2+}$ ions. A photodissociation laser is used to dissociate formed molecules of $\mathrm{Th}^{+}$. The lasers used for the spectroscopy of $\mathrm{Th}^{2+}$ are shown on the left side of the optical setup.}
    \label{Fig:1}
\end{figure}

\begin{figure}[h]
    \centering
    \includegraphics[width=0.48\textwidth]{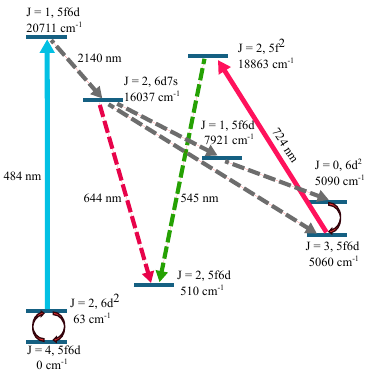}
    \caption{Level scheme that is used to populate the $J = 0$ state in $\mathrm{Th}^{2+}$. Solid arrows indicate applied lasers, dashed arrows indicate spontaneous decay from ions in different states, and curved arrows indicate collisional mixing.}
    \label{Fig:2}
\end{figure}

The apparatus for trapping $\mathrm{Th}^{2+}$ ions used in this experiment is the same as described in Ref. \cite{PhysRevA.85.033402,Thielking2018}. We first load $\mathrm{Th}^{+}$ ions by laser ablation from $^{232}\mathrm{Th}$ and $^{229}\mathrm{Th}$ targets. The ions are cooled to near room temperature by collisions with buffer gas (Ar or He). The optical setup with all the lasers used is shown in Fig.~\ref{Fig:1}.
After loading $\mathrm{Th}^{+}$, we ionize it to $\mathrm{Th}^{2+}$ via a three-photon process, which includes one photon of 402~nm and two photons from the photoionization laser at 269~nm \cite{PhysRevA.88.012512}. The rate of photoionization is controlled by adjusting the 269~nm laser power to maintain a constant number of $\mathrm{Th}^{2+}$ ions throughout the experiment. For laser excitation towards the $5090_{0}$ level, we start with ions from the $63_{2}$ state that is populated with about 40\% of the ions by collisions with the buffer gas. The level scheme of the transitions used for excitation and fluorescence detection is shown in Fig.~\ref{Fig:2}. Collisional mixing is illustrated by curved arrows. It generates the dominant decay mechanism for the metastable level in our experiment. The experimental sequence runs at a rate of $1\,\mathrm{kHz}$. To populate the $5090_{0}$ state, we apply a 484 nm laser pulse for $150\,\mu\mathrm{s}$ duration that drives the transition $63_{2} \rightarrow 20711_{1}$. The decay of the $20711_{1}$ level populates a detectable number of ions in the $5090_{0}$ state via the channel shown in Fig.~\ref{Fig:2}.
Increasing the pulse duration of the 484~nm laser does not improve the signal, as the lifetime of the metastable state in the buffer gas is of the same order. After $150\,\mu\mathrm{s}$, the 484~nm laser is turned off, and fluorescence from the metastable state is detected to verify the population there. The detailed scheme and the energy levels used are described in the next section. For the lifetime measurement, the experiment is performed using the $^{232}\mathrm{Th}$ isotope because of easier loading of ions. 
We continuously monitor the fluorescence through the $5060_{3}$ state, similar to the optical shelving method \cite{PhysRevA.69.032503, PhysRevA.105.042801}. This method allows us to continuously monitor the decay of ions from the metastable state through collisional mixing. As the lifetime is less than $1\,\mathrm{ms}$ in our setup due to collisional mixing, we average the data at a rate of $1\,\mathrm{kHz}$. At the end of the sequence, we apply a laser pulse to monitor the number of $\mathrm{Th}^{+}$ ions, along with an ionization pulse and a pulse at 266~nm to dissociate 
molecular ions that $\mathrm{Th}^{+}$ may have formed in reactions with the background gas.

\section{Detection of population in the metastable state}
\begin{figure}[h]
    \centering
    \includegraphics[width=0.48\textwidth]{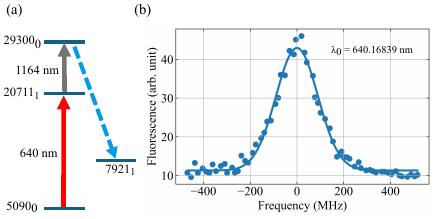}
    \caption{Fluorescence detection after two-photon excitation from the metastable state of $^{232}\mathrm{Th}^{2+}$. Solid markers represent experimental data, and the line corresponds to the fitted curve.}
    \label{Fig:3}
\end{figure}

To detect ions present in the $5090_{0}$ state, we use a two-photon excitation scheme, coupling $5090_{0} \rightarrow 20711_{1} \rightarrow 29300_{0}$ states. This scheme is shown in Fig.~\hyperref[Fig:3]{\ref*{Fig:3}a} with the corresponding energy levels and their $J$ values. The previously used two-photon scheme from $63_{2} \rightarrow 20711_{1} \rightarrow 29300_{0}$ enabled us to identify the wavelength for the transition $20711_{1} \rightarrow 29300_{0}$, which corresponds to a wavelength of 1164.3~nm \cite{Thielking2018}. We stabilize the 1164~nm laser and scan the 640~nm laser to couple $5090_{0} \rightarrow 20711_{1}$. In Fig.~\hyperref[Fig:3]{\ref*{Fig:3}b}, we show the detected fluorescence signal that is proportional to the population of the metastable state for $^{232}\mathrm{Th}^{2+}$. The measured full width at half maximum (FWHM) is 210~MHz, determined by the frequency selectivity of the second step. The value for the transition wavelength reported in Ref. \cite{osti_6348879, Biémont_2002} for the 640~nm transition lies within this range, indicating good agreement for the position of the $5090_0$ level for $^{232}\mathrm{Th}^{2+}$.

\begin{figure}[h]
    \centering
    \includegraphics[width=0.48\textwidth]{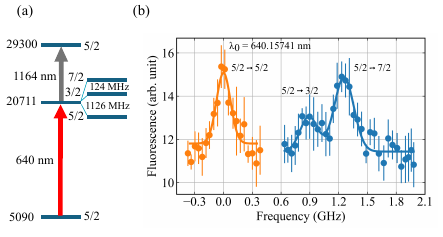}
    \caption{Fluorescence detection after two-photon excitation from the metastable state of $^{229}\mathrm{Th}^{2+}$.  The wavelength is given for the $5/2 \rightarrow 5/2$. $F \rightarrow F$ labels are shown for the 640~nm transition. Solid markers represent experimental data with error bars, and the lines correspond to the fitted curve.}
    \label{Fig:4}
\end{figure}

For $^{229}\mathrm{Th}$, with a ground state nuclear spin of $I = 5/2$, the hyperfine structure gives rise to three $F$ levels in the intermediate state ($J = 1$). The scheme and energy levels splitting are shown in Fig.~\hyperref[Fig:4]{\ref*{Fig:4}a}. The 1164~nm laser is first locked to the $F=5/2 \rightarrow 5/2$ transition, while the 640~nm laser is scanned to determine the exact wavelength for $^{229}\mathrm{Th}^{2+}$. A fluorescence signal is observed as shown in Fig.~\hyperref[Fig:4]{\ref*{Fig:4}b}, yellow color corresponds to the $F = 5/2 \rightarrow 5/2 \rightarrow 5/2$ transition. The FWHM of this transition is approximately 175~MHz.

To observe the other two intermediate levels which lie within the Doppler broadening of the 640~nm transition, the 1164~nm laser is locked to the $7/2 \rightarrow 5/2$ transition, while the 640~nm laser is scanned. The strong observed peak (shown in blue in Fig.~\hyperref[Fig:4]{\ref*{Fig:4}b}) corresponds to the $5/2 \rightarrow 7/2 \rightarrow 5/2$ transition, whereas the weak peak corresponds to the $5/2 \rightarrow 3/2 \rightarrow 5/2$ transition. The measured frequency separation for 640~nm between the $5/2 \rightarrow 5/2$ and $5/2 \rightarrow 7/2$ transitions is approximately equal to the applied shift of the 1164~nm laser, that is, 1250~MHz. The frequency difference between the $5/2 \rightarrow 7/2$ and $5/2 \rightarrow 3/2$ transitions is influenced by the choice of the 1164~nm laser frequency, as both transitions lie within the Doppler broadening.
When the 1164~nm laser is locked to the $7/2 \rightarrow 5/2$ transition, the $5/2 \rightarrow 3/2$ transition experiences an extra frequency shift given by $\Delta f_1 = \Delta f_2 \cdot (1+k_1/k_2)$ in the co-propagating geometry \cite{PhysRevA.14.751, PhysRevA.101.042503}, which corresponds to 350~MHz for $\Delta f_2$ of 124~MHz. Here, we also observed a peak shift of approximately 350~MHz between $5/2 \rightarrow 3/2$ and $5/2 \rightarrow 7/2$.

\subsection{Measurement of the isotope shift of the metastable state in $^{232}\mathrm{Th}^{2+}$ and $^{229}\mathrm{Th}^{2+}$}
As a measure of the interaction with the nucleus, we determine the isotope shift $(\nu_{229} - \nu_{232})$ for the forbidden transition 
$63_{2} \rightarrow 5090_{0}$. We start from the known isotope shift for the $63_{2} \rightarrow 20711_{1}$ transition, which is 8.2(2)~GHz \cite{PhysRevLett.121.213001}. The hyperfine constants $A$ and $B$ for the level $20711_{1}$ are taken from Ref. \cite{PhysRevA.98.020503}. Using these constants, we calculated the position of the centroid frequency relative to the peak $5/2 \rightarrow 5/2$ for the $5090_{0} \rightarrow 20711_{1}$ transition for $^{229}\mathrm{Th}^{2+}$. The centroid is found to be shifted by approximately 800~MHz relative to the $5/2 \rightarrow 5/2$ peak shown in Fig.~\hyperref[Fig:4]{\ref*{Fig:4}b}. From this, the total isotope shift for the $5090_{0} \rightarrow 20711_{1}$ transition between $^{232}\mathrm{Th}^{2+}$ and $^{229}\mathrm{Th}^{2+}$ is 8.8(2)~GHz at 640~nm. Knowing that the isotope shift for the $63_{2} \rightarrow 20711_{1}$ transition is 8.2(2)~GHz, the resulting isotope shift for the $63_{2} \rightarrow 5090_{0}$ transition is therefore 0.6(3)~GHz. This reduced isotope shift arises from the nearly identical electronic configurations of the $6d^2\,{}^3F_2\,(63\ \mathrm{cm^{-1}})$ and $6d^2\,{}^3P_0\,(5090\ \mathrm{cm^{-1}})$ states, and from the small field shift associated with the d-electron orbitals in $\mathrm{Th}^{2+}$ \cite{PhysRevLett.121.213001,PhysRevA.98.020503}. 

\section{Lifetime measurements}
\begin{figure}[h]
    \centering
    \includegraphics[width=0.48\textwidth]{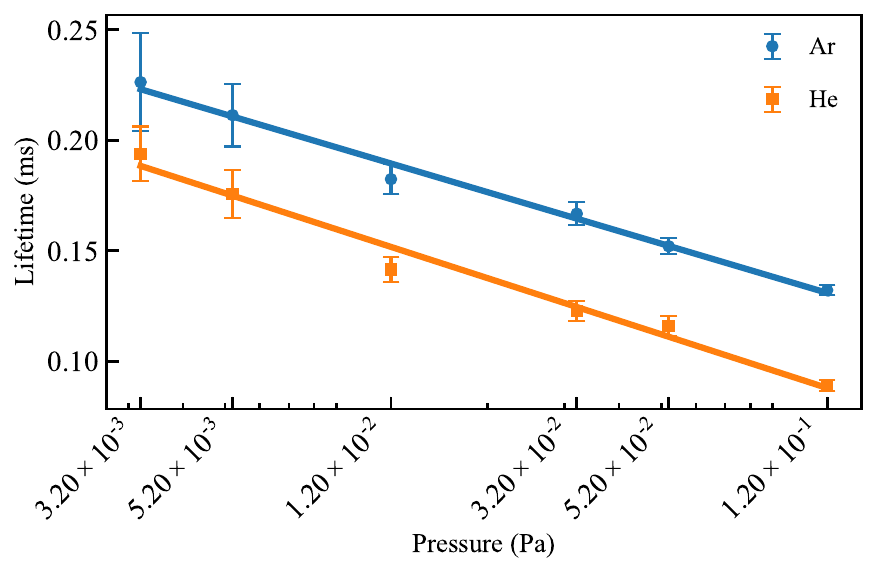}
    \caption{Lifetime measurement of the $J = 0$ state at different pressures for the buffer gases Ar and He. The lifetime corresponds to the time constant for slow decay ($\tau_s$) as seen in Fig. 5.}
    \label{Fig:6}
\end{figure}

In our experiment, the lifetime of the $5090_0$ state is determined by collisional mixing with the nearby $5060_{3}$ level. The $5060_{3}$ state decays to the $63_{2}$ level through an electric dipole transition. We monitor the population of the $5060_{3}$ state in fluorescence at 545 nm by exciting a transition with a wavelength of 724~nm to the $18863_{2}$ level as shown in Fig.~\ref{Fig:2}. A pulse of $150\,\mu\mathrm{s}$ duration of the 484~nm laser drives the $63_{2} \rightarrow 20711_{1}$ transition, and decay of the $20711_{1}$ state populates the $5090_{0}$ and $5060_{3}$ levels. The 724~nm laser is applied continuously. 
\begin{figure}[h]
    \centering
    \begin{subfigure}[t]{0.48\textwidth}
        \centering
        \caption{} 
        \includegraphics[width=\textwidth]{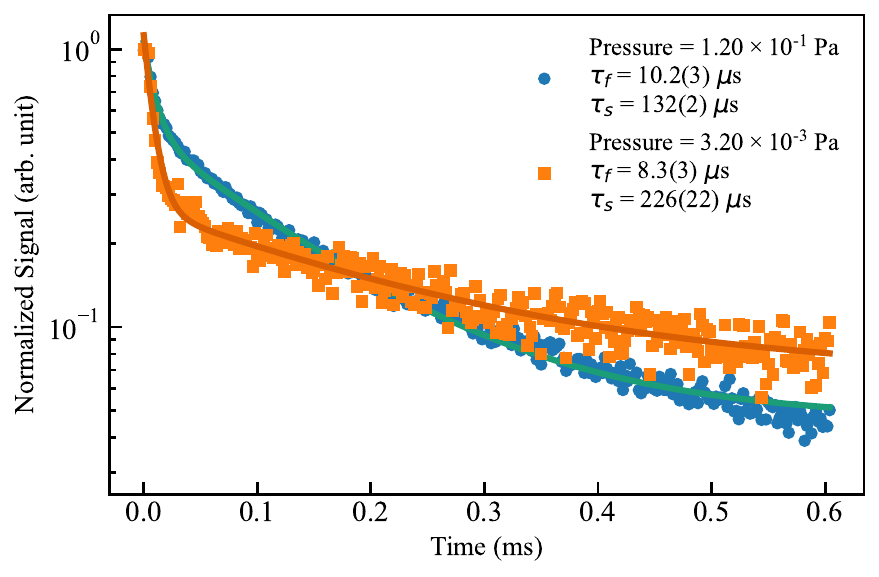}
        \label{Fig:5a}
    \end{subfigure}
    \begin{subfigure}[t]{0.48\textwidth}
        \centering
        \caption{}
        \includegraphics[width=\textwidth]{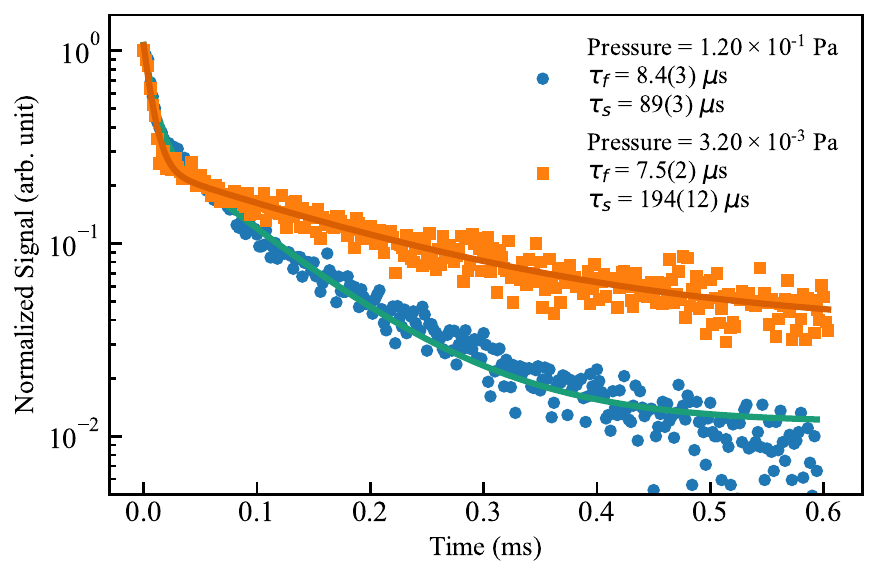}
        \label{Fig:5b}
    \end{subfigure}
    \caption{(a) Decay of the $J = 0$ state in the presence of Ar as buffer gas at two different pressures. The data are fitted with a bi-exponential function. Here, $\tau_f$ and $\tau_s$ correspond to the time constants of the fast and slow decay components. (b) Decay in the presence of He as buffer gas at two different pressures.}
    \label{Fig:5}
\end{figure}
After turning off the 484~nm pumping laser, the fluorescence signal as shown in Fig.~\ref{Fig:5} exhibits an initial fast decay followed by a slower decay. The fast decay corresponds to the loss of initial population from $5060_{3}$, while the slow decay is directly connected to the lifetime of the $5090_{0}$ level which populates the $5060_{3}$ level through collisional mixing. In Fig.~\ref{Fig:5a}, the data is taken with argon as a buffer gas at two different pressures. We fit the data with a bi-exponential fit to extract the time constants for fast and slow decay. The time constant for slow decay is determined by the rate of collisional mixing and limits the lifetime in our setup. A detailed study of multi-exponential decay is presented in Refs. \cite{PhysRevA.58.1111, Trabert2000, atoms12120073}. Detecting the lifetime through the population of $5060_{3}$ allows us to continuously record the fluorescence signal using a scalar card with a time resolution of 100~ns and an averaging rate of 1 kHz. This method is not prone to fluctuations in the initial ion number over longer timescales, which occur in our setup due to molecule formation of $\mathrm{Th}^{2+}$. Similarly, we also performed the measurement with helium as the buffer gas. The result is shown in Fig.~\ref{Fig:5b}.
To understand the behavior of the mixing, we performed measurements at different pressures. A linear decrease in lifetime with $\log(P)$ is observed as shown in Fig.~\ref{Fig:6}.
For argon as the buffer gas, collisional mixing is slightly slower than for helium, resulting in a longer lifetime.
A similar difference in collision dynamics has been observed in \cite{Thielking2018}. 
The longest observed lifetime of the $5090_{0}$ state is significantly shorter than the calculated radiative lifetime of the level of approximately 95~s \cite{PhysRevA.90.032512}. Our measurement  
cannot be directly extrapolated to ultrahigh vacuum conditions although the limit arises from collisional mixing \cite{PhysRevA.99.022706, Endo:21}. Reducing the buffer gas pressure in our setup has the side-effect of increasing the temperature of the ions to above room temperature 
because of radiofrequency heating in the ion trap.

\section{Conclusions}
In this experiment, we presented studies on the metastable state with $J = 0$ in $\mathrm{Th}^{2+}$, demonstrating schemes to efficiently populate the state by laser excitation at 484~nm and to detect it in fluorescence after excitation at 724~nm. We determined an isotope shift of $0.6(3)\ \mathrm{GHz}$ between $^{232}\mathrm{Th}^{2+}$ and $^{229}\mathrm{Th}^{2+}$, smaller by an order of magnitude than previously reported isotope shifts for other transitions in $\mathrm{Th}^{2+}$. The small isotope shift suggests that the $5090_{0}$ state 
has minimal interaction with the nuclear charge distribution and identical field shift coefficient as $63_{2}$ \cite{PhysRevA.92.020503, PhysRevLett.121.213001}. We also measured the lifetime at different pressures and observed a maximum lifetime of 225~\textmu s at $3.2\times10^{-3}$~Pa of argon buffer gas. In the present setup, the lifetime is limited by collisional mixing, which would be strongly suppressed for cooled ions in ultrahigh vacuum.

\begin{figure}[]
    \centering
    \includegraphics[width=0.48\textwidth]{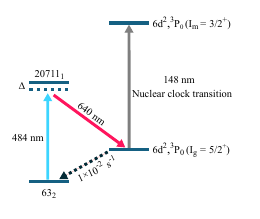}
    \caption{Schematic of a hyperfine-free nuclear clock with Th$^{2+}$. Here 484~nm and 640~nm lasers drive a coherent Raman transition for state preparation. $\Delta$ is the detuning used to suppress off-resonant scattering.}
    \label{Fig:7}
\end{figure}

As an outlook, we propose a hyperfine-free nuclear optical clock by preparing $^{229}\mathrm{Th}^{2+}$ in the long-lived $5090_{0}$ state as shown in Fig.~\ref{Fig:7}. The long lifetime ($\approx 95~\mathrm{s}$) of the $5090_{0}$ state allows long interrogation times of the nuclear transition, whereas the $s$-state with $J = 1/2$ in $\mathrm{Th}^{3+}$ is limited to a lifetime of only $\sim 600~\mathrm{ms}$ \cite{PhysRevA.86.060501,E.Peik_2003}. 
A minor drawback of using $J=0$ instead of $J=1/2$ is that the nuclear transition will be split into Zeeman components with first-order magnetic field sensitivities between $\pm0.4$ and $\pm6$ Hz/$\mu$T. A potential systematic frequency offset can be eliminated by averaging over two Zeeman components with opposite signs of the shift.

A viable experimental setup will consist of sympathetically cooled $\mathrm{Th}^{2+}$ with co-trapped laser cooled ions as it is done for $\mathrm{Th}^{3+}$ \cite{PhysRevA.109.033116, Moritz2025}. Sr$^+$ or In$^+$ would be suitable  species because their charge-to-mass ratio is similar to that of Th$^{2+}$. To prepare ions in $5090_{0}$, a stimulated Raman scheme can be used to transfer the ions from $63_{2}$ to $5090_{0}$ via $20711_{1}$ as an intermediate state, also shown in Fig.~\ref{Fig:7}. The lasers at 484~nm and 640~nm, as used in this work, can enable coherent population transfer through the Raman process with a few milliwatts of optical power \cite{spahn2025demonstrationramanvelocityfilter, qk3v-46y8, PhysRevA.105.032618}. Since no electronic levels of Th$^{2+}$ are known in the vicinity of 9.0~eV excitation energy, applying laser radiation at 148~nm to drive the nuclear transition from the $5090_{0}$ level is not expected to lead to parasitic electronic excitation. The detection of nuclear excitation can be performed with quantum logic spectroscopy via the same co-trapped ions used for sympathetic cooling \cite{schmidt2005spectroscopy, Micke2020}.

Studying the 8.4~eV nuclear resonance of $^{229}$Th in different electronic configurations of atomic ions will provide an opportunity to compare nuclear-electronic hyperfine interactions for a variety of cases. Examples of strong coupling include 
electronic bridge processes in Th$^+$ \cite{Shigekawa2026} and in Th$^{35+}$ \cite{PhysRevLett.124.192502, Porsev_2021}, and hyperfine-mixing in hydrogen-like Th$^{89+}$ \cite{PhysRevC.57.3085,PhysRevC.64.064301}.
In contrast, the $J=0$ states in Th$^{4+}$ \cite{zcsk-3lw5} and Th$^{2+}$ as discussed here are cases where the leading hyperfine interactions of the nucleus with the electron shell are minimized.

\section{Acknowledgments}
We would like to thank N. Huntemann for providing the accurate wavelength reference at 871~nm used to calibrate the wavemeter. We also thank G. Zitzer and N. Irwin for helpful discussions and PTB Department 5.5 and S. Hennig for the manufacturing and preparation of the Th ablation targets. This work has been funded by the European Research Council (ERC) under the European Union’s Horizon 2020 research and innovation programme (Grant Agreement No. 856415), by Deutsche Forschungsgemeinschaft (DFG) – SFB 1227 - Project-ID 274200144 (Project B04), and by the Max-Planck-RIKEN-PTB-Center for Time, Constants and Fundamental Symmetries. The project 23FUN03 HIOC [Grant DOI: 10.13039/100019599] has received funding from the European Partnership on Metrology, co-financed from the European Union’s Horizon Europe Research and Innovation Program and by the Participating States. 
\bibliographystyle{apsrev4-2}
\bibliography{Ref}
\end{document}